&figures not included

\documentstyle[12pt]{article}
\input{preprint}

\begin{document}
\begin{center}
Coulomb drag between parallel two-dimensional electron systems

by

Antti-Pekka Jauho$^{\dagger}$

Nordita, Blegdamsvej 17, Copenhagen 2100  {\O}, Denmark

\vspace{0.5cm}

 and

 Henrik Smith

Physics Laboratory, H. C. {\O}rsted Institute,

Universitetsparken 5, Copenhagen 2100 {\O}, Denmark

\vspace{1cm}

ABSTRACT

\end{center}

\noindent { The Coulomb contribution to
  the  temperature-dependent rate of momentum transfer, $1/\tau_D$,
between  two electron systems in  parallel layers is determined
by setting up  two coupled Boltzmann equations,
 with the boundary condition that no current flows in the layer where
an induced voltage is measured.
The effective Coulomb interaction between the layers
is determined selfconsistently, allowing for the finite  thickness
of the layers.
As $T\rightarrow 0$, we find that $1/\tau_DT^2$ approaches a
constant value. At  higher temperatures $1/\tau_DT^2$
exhibits a maximum at  $T=T_{\rm max}$ and then
 decreases as $1/T$ with increasing
temperature. The value of $T_{\rm max}$ depends on the
layer separation $d$ according to $T_{\rm max}\propto d^{-\alpha}$,
where $\alpha\simeq 0.8$. The overall magnitude of the
calculated $1/\tau_D$ is approximately one half
of the results of a recent experiment, suggesting that
other mechanisms of  momentum transfer may be important.}

\vspace{.5cm}
PACS numbers: 72.10-d, 72.20.Dp, 73.50.Dn
\vspace{.5cm}

1. {\sc INTRODUCTION}

It is well known that the
 electron-electron scattering rate $1/\tau(\epsilon)$ in a
 three-dimensional electron gas at zero
temperature depends on the
 electron energy $\epsilon$ according to
 $1/\tau(\epsilon)\propto (\epsilon-\mu)^2$,
where $\mu$ is the chemical potential \cite{lan}. At finite temperatures
this characteristic energy dependence
 yields relaxation rates that are proportional to $T^2$.
The effect of electron-electron scattering on the
transport properties of ordinary metals is usually  weak
compared to the competing effects of electron-impurity  and electron-phonon
scattering. The
$T^2$-dependence associated with electron-electron scattering
has been observed in metals at relatively high temperatures
by measuring  deviations from the Wiedemann-Franz-Lorenz law (for
a discussion of this and other consequences
of electron-electron scattering, see e.\ g.\ \cite{sj}).
At low tem\-pera\-tures the effects of electron-electron scattering
on  various transport properties are often
 difficult to separate from the  effects of other inelastic
processes. The phase-breaking due to electron-electron scattering  has
important consequences for the localization phenomena
occurring in disordered systems and it has been
investigated extensively in this context \cite{lee}.

The characteristic energy-dependence of the electron-electron scattering rate
is due to the phase-space restrictions that apply to the mutual scattering
of particles in a nearly degenerate gas.
These restrictions  are different in two and in three
dimensions. For a two-dimensional electron gas one finds that
the  scattering rate at zero temperature is proportional
\linebreak
to
$(\epsilon-\mu)^2\ln|\epsilon-\mu|$, as first
 shown by Hodges et al.\ \cite{hod}
and, independently, by Chaplik \cite{chap}. At finite temperatures
the corresponding relaxation rates become proportional to
$T^2\ln T$. The non-analytic temperature-dependence
can be traced to a logarithmic
singularity for small values of the momentum transfer involved
in a scattering process.
The logarithmic  energy-dependence
 $(\epsilon-\mu)^2\ln|\epsilon-\mu|$ was recently confirmed experimentally
 \cite{yac} by measurements of quantum interference in a
two-dimensional electron gas.

In  a  recent experiment
  Gramila et al.\ \cite{gram} measured the mutual friction between
two parallel two-dimensional electron systems as a  function
of  the temperature and the distance between the two layers.
 The systems investigated consisted of modulation-doped
 GaAs/AlGaAs double quantum wells,  grown
by molecular-beam epitaxy. The observed frictional drag
 was interpreted
as being due  to the Coulomb interaction between the
two separate electron systems.
 Gramila et al.\ \cite{gram} also performed a calculation
of the drag in the low temperature limit,
and found the
rate of momentum transfer   to be proportional to
 $T^2$. Although
the  overall temperature dependence  of their
calculated rate was in qualitative agreement with
 experiment,   there remained significant differences,
in particular for samples with large values of the layer
separation \cite{gram1}.

The purpose of the present paper
is to investigate the Coulomb drag problem
in   detail by calculating the rate of momentum
transfer  as a function
of temperature. In the limit
of very low temperatures  we find that the rate of momentum transfer
is proportional to $T^2$, in agreement with
the result obtained in ref.\ \cite{gram}. At
temperatures which are somewhat higher
(but still small compared to the Fermi temperature $T_{\rm F}$)
 our calculated
temperature dependence differs from $T^2$. As we shall see, the deviation
from the $T^2$-dependence is in qualitative, but not quantitative
agreement with the one observed experimentally in ref.\ \cite{gram}, where
 the momentum relaxation rate divided by
$T^2$ was found to exhibit a maximum
around $T=2$ K, which is about thirty times less
than  $T_{\rm F}$.

Although the indirect electron-electron
interaction caused by the  virtual exchange
of phonons may  also affect  the observed friction \cite{gram}
 \cite{gram1},
the scope of the present work
is limited to evaluating the temperature-dependence
arising from the  Coulomb interaction and comparing the calculated
relaxation rate with the one measured experimentally.
Our formulation is, however, sufficiently general that
other coupling mechanisms besides the direct Coulomb interaction
may  be incorporated.

The possibility of observing the Coulomb  drag effect
in heterostructures was  suggested by Price \cite{Price}.
A theoretical formulation      of the drag problem
was given by Pogrebinskii \cite{Pog}.
The effect has been  considered
in both one-, two- and three-dimensional systems
by a number of different authors, cf. refs.
 \cite{Laik}, \cite{Laik1}, \cite{Boiko}, \cite{Rojo},
\cite{Maslov} and \cite {Tso}. Most of these
calculations treat
only high- or low-temperature limits,
and to our knowledge the full temperature dependence
of the  Coulomb drag has not been considered before.
Momentum transfer between
two coupled electron-hole gases was studied
recently by Sivan et al.\ \cite{sivan}.

The plan of the paper is as follows: In sect.\ 2
we define the drag resistivity and the associated
momentum relaxation rate in terms of experimentally
observable quantities. The theoretical model
used for calculating the momentum relaxation rate
is treated in sect.\ 3, where we    set up the
two coupled Boltzmann equations and derive a  general
expression for  the momentum relaxation rate. We also
  discuss  the screening of
the Coulomb interaction between electrons in the two
layers.  The resulting effective interaction enters
our expression for the momentum relaxation rate,
which  in general must be evaluated
numerically. In sect.\ 4 we perform an analytic evaluation of
the rate of momentum relaxation at very low temperatures
and exhibit the results of the numerical calculation at
arbitrary temperatures. The theoretical results are compared to
the measured momentum relaxation rate,  in regard to
its absolute magnitude as well as
 its dependence on temperature and layer separation.
 The  detailed derivation of the screened Coulomb interaction
is deferred to
App.\ A, and technical details
of an alternative integration procedure can be found in App.\ B.

\vspace{1cm}

2. {\sc DRAG RESISTIVITY}

Let us consider two parallel layers each containing a two-dimensional gas
of electrons, as illustrated in Fig.\ 1. The perpendicular distance
between the midpoints of the
two layers is denoted by $d$. In the experiment performed by
Gramila et al.\ \cite{gram} a current $I_2$ is driven along layer 2,
and one measures the voltage difference $V_1$ which is induced
in layer 1 under the condition that no current flows in this  layer.
 The current per unit width, $j_2$, in layer 2
is given by $j_2=I_2/W$, while the magnitude of the electric field
strength, which prevents the electrons in layer 1 from
being dragged along by the current in
layer 2, is $E_1=V_1/l$. Here $W$ is the width of
layer 2, while $l$ denotes the distance between the potential
probes shown in Fig.\ 1. The  drag resistivity $\rho_D$ is then defined by
\begin{equation}
\rho_D=\frac{E_1}{j_2}=\frac{WV_1}{I_2l}.
\end{equation}
Following ref.\ \cite{gram}
it is convenient to translate the drag resistivity into a
momentum relaxation rate $1/\tau_D$ by defining the
drift velocity ${ u}_2$ according to
\begin{equation}
{ j}_2=n_2e{ u}_2,
\label{2}
\end{equation}
where $n_2$ is the number  of electrons per unit area
in layer 2, while $e$
is the elementary charge.
Experimentally, the  electric field $E_1$ is
found to be proportional to $u_2$. The coefficient relating the
two quantities is the  drag mobility $\mu_D$, which is defined by
\begin{equation}
\mu_D=\frac{u_2}{E_1}.
\label{3}
\end{equation}
The mobility $\mu_D$ may in turn be expressed in terms of
the momentum relaxation rate $1/\tau_D$, defined according to
\begin{equation}
\mu_D=\frac{e}{m}\tau_D,
\label{4}
\end{equation}
where $m$ is the effective mass of the conduction electrons (in the
case of AlGaAs/GaAs, the effective mass $m$ is
0.067 times the electron mass $m_e$).
The drag resistivity $\rho_D$, which has the  dimension of a
resistance, may thus be written as
\begin{equation}
\rho_D=\frac{m}{n_2e^2\tau_D}.
\label{drag}
\end{equation}

In summary, the ratio between the observed voltage $V_1$ and
the imposed current $I_2$ is expressed in terms
of  the drag resistivity $\rho_D$ by the definition
\begin{equation}
\frac{V_1}{I_2}=\frac{l}{W} \rho_D,
\label{def}
\end{equation}
where $\rho_D$  according to (\ref{drag})
may be written in terms of the momentum relaxation rate $1/\tau_D$.
The equations (\ref{drag}) and (\ref{def}) are
 simply a definition of a convenient
quantity $\tau_D$, which we  determine
in the following section as a function
of the temperature $T$ and the distance $d$ separating the two layers,
starting from  the Boltzmann equation for the distribution function of the
electrons.

\vspace{1cm}

3. {\sc THE MOMENTUM RELAXATION RATE}

The  momentum relaxation rate is determined
by using  the linearized Boltzmann equation
to derive a balance equation
between the induced electric field and
the drag due to the drive current.
 In the presence of an electric field
${\bf E}_1$ this linearized Boltzmann equation is
\begin{equation}
\dot{{\bf k}}_1\cdot\frac{\partial f^0}{\partial {\bf k}_1} =
(\frac{
\partial f_1}{\partial t})_{\rm coll},
\label{bol}
\end{equation}
where
\begin{equation}
\hbar\dot{{\bf k}}_1=-e{\bf E}_1,
\label{motion}
\end{equation}
while $f^0$ is the equilibrium distribution function
 (we label quantities
referring to layer 1 by 1, $1'$ etc. and
similarly for layer 2).
The linearized collision integral is
\begin{eqnarray}
(\frac{
\partial f_1}{\partial t})_{\rm coll}&=&
-\sum_{\sigma_2\sigma_{1'}\sigma_{2'}}^{}
\int_{}^{}\frac{d{\bf k}_2}{(2\pi)^2}
\int_{}^{}\frac{d{\bf k}_{1'}}{(2\pi)^2}w(1,2;1',2') \nonumber \\
& & (\psi_1+\psi_2-\psi_{1'}-\psi_{2'})f^0_1f^0_2(1-f^0_{1'})(1-f^0_{2'})
\delta(\epsilon_{1} + \epsilon_{2} - \epsilon_{1'} - \epsilon_{2'}).
\label{collint}
\end{eqnarray}
Here the function $w(1,2;1',2')$  determines the probability
that two electrons in states
 ${\bf k}_1 \sigma_1$ and
 ${\bf k}_2 \sigma_2$ will scatter to
 ${\bf k}_{1'} \sigma_{1'}$ and
 ${\bf k}_{2'} \sigma_{2'}$.
In the Born approximation $w$ is proportional to
the square of the Fourier-transform of the effective interaction
as specified below.
The deviation function $\psi$  has been introduced by the  definition
\begin{equation}
f-f^0=f^0(1-f^0)\psi,
\end{equation}
implying that $\psi$ vanishes in equilibrium.

The current flowing in layer 2 is assumed to be limited by
 impurity scattering, corresponding to the deviation function
\begin{equation}
\psi_2=-\frac{1}{kT}\tau_2ev_{2x}E_2,
\label{noneq}
\end{equation}
where ${\bf E}_2$ is the electric field  in layer 2, directed along the
 $x$-axis, and $\tau_2$ is an (energy-independent)
momentum relaxation time, which
 determines the electron mobility $\mu_2$ according to $\mu_2=e\tau_2/m$.
 Since no current is flowing in layer 1, the electron
distribution in this layer
is taken to be the equilibrium one, corresponding to $\psi_1 =\psi_{1'}=0$.
According to (\ref{bol})-(\ref{collint}) the electric field ${\bf E}_1$
  then balances the drag resulting from
 the fact that $\psi_2$ and $\psi_{2'}$ are non-zero.
 In order to determine $E_1$ we insert the
non-equilibrium distribution function given by (\ref{noneq}) into
(\ref{collint}). Due to momentum conservation we have
\begin{equation}
v_{2x}-v_{2'x}=v_{1'x}-v_{1x},
\end{equation}
since the effective masses $m$ are assumed to be identical
in the two layers.
We now multiply both sides of (\ref{bol}) by $k_{1x}$ and sum
over the states ${\bf k}_1\sigma_1$. This yields
\begin{equation}
2\int_{}^{}\frac{d{\bf k}_1}{(2\pi)^2}k_{1x}\frac{(-e)E_{1}}{\hbar}
\frac{\partial}{\partial k_{1x}}f^0(\epsilon_1) =
\sum_{\sigma_1}^{}\int_{}^{}\frac{d{\bf k}_1}{(2\pi)^2}k_{1x}(\frac{
\partial f_1}{\partial t})_{\rm coll}.
\label{boleq}
\end{equation}
 The left hand side of (\ref{boleq})
then becomes $eE_1n_1/\hbar$ after
 use of partial integration, with $n_1$ being  the electron density,
which is related
to the Fermi wave-vector  $k_{\rm F}$ according to
\begin{equation}
n_1=\frac{k_{\rm F}^2}{2\pi}.
\end{equation}
 We have assumed throughout this paper that the electron number density
is the same in the two layers, $n_1=n_2$. It is
straightforward to generalize our treatment
 to the case where the two densities differ.

The right hand side of (\ref{boleq})
may be simplified, as the quantity
$k_{1x}(k_{1'x}-k_{1x})$ in the integrand
 may be replaced by
 $-(k_{1'x}-k_{1x})^2/2$, because of the symmetry of the
remaining part of the integrand with
respect to the interchange of 1 and $1'$.
Furthermore, since the calculated momentum relaxation rate is
 independent of whether the electric field is  taken to
be  along the $x$- or
the $y$-axis, we   add the corresponding contributions and
divide by 2, resulting in the replacement of
$(k_{1'x}-k_{1x})^2/2$ by $q^2/4$,
where ${\bf q}$ is the wave-vector transfer given by
\begin{equation}
{\bf q}={\bf k}_{1'}- {\bf k}_{1}.
\end{equation}
 This allows
us to further simplify the right hand side of (\ref{boleq}):
\begin{eqnarray}
\sum_{\sigma_1}^{}\int_{}^{}\frac{d{\bf k}_1}{(2\pi)^2}k_{1x}(\frac{
\partial f_1}{\partial t})_{\rm coll}&=&
-\frac{e\hbar E_2\tau_2}{4mkT}\sum_{\sigma_1\sigma_2\sigma_{1'}\sigma_{2'}}^{}
       \int_{}^{}\frac{d{\bf k}_1}{(2\pi)^2}
\int_{}^{}\frac{d{\bf k}_2}{(2\pi)^2}
\int_{}^{}\frac{d{\bf k}_{1'}}{(2\pi)^2}w(1,2;1',2') \nonumber \\
& & q^2  f^0_1f^0_2(1-f^0_{1'})(1-f^0_{2'})
\delta(\epsilon_{1} + \epsilon_{2} - \epsilon_{1'} - \epsilon_{2'}).
\label{lign1}
\end{eqnarray}
There are two ways to proceed from
(\ref{lign1}). First, one may use the delta-function to carry
out the integration over the angle between
${\bf k}_1$ and ${\bf k}_2$, as explained in App.\ B. Here in the main text
we shall make the simplifying assumption that
 $w$  only depends on ${\bf q}$ (and possibly
$\epsilon_1-\epsilon_{1'}$). Then we introduce
${\bf q}$ as an integration variable in (\ref{lign1}). It is
convenient  \cite{Quinn}
 to express (\ref{lign1}) in terms of the
two-dimensional susceptibility function $\chi(q,\omega)$, defined by
\begin{equation}
\chi({ q}, \omega)
=-\int_{}^{}\frac{d{\bf k}_1}{(2\pi)^2}\frac{f^0(\epsilon_{1})
-f^0(\epsilon_{1'})}{\epsilon_{1}
-\epsilon_{1'}+\hbar \omega +i\delta}.
\end{equation}
The imaginary part of the susceptibility, ${\rm Im}\chi$,
is given at zero temperature by
\begin{eqnarray}
{\rm Im}\chi(q,\omega)&=&\frac{mk_{\rm F}}{2\pi\hbar^2 q}\left\{
\sqrt{1-(\frac{\omega}{v_{\rm F}q}
-\frac{q}{2k_{\rm F}})^2} -
\sqrt{1-(\frac{\omega}{v_{\rm F}q}
+\frac{q}{2k_{\rm F}})^2}\,\right\}\nonumber \\
& & {\rm for} \;\,q<2k_{\rm F},\;\;
0<\omega<v_{\rm F}q(1-q/2k_{\rm F})\nonumber \\
   &=&\frac{mk_{\rm F}}{2\pi\hbar^2 q} \sqrt{1-(\frac{\omega}{v_{\rm F}q}
-\frac{q}{2k_{\rm F}})^2}\nonumber \\
& & {\rm for}\;\,q<2k_{\rm F},\;\;
v_{\rm F}q(1-q/2k_{\rm F})<
\omega<v_{\rm F}q(1+q/2k_{\rm F}), \nonumber \\
   &=&\frac{mk_{\rm F}}{2\pi\hbar^2 q}\sqrt{1-(\frac{\omega}{v_{\rm F}q}
-\frac{q}{2k_{\rm F}})^2}\nonumber \\
& &  {\rm for}\;\,q>2k_{\rm F},\;\;
v_{\rm F}q(-1+q/2k_{\rm F})<
\omega<v_{\rm F}q(1+q/2k_{\rm F}), \nonumber \\
   &=&0, \;\;{\rm otherwise}.
\end{eqnarray}
Strictly speaking we should here use the
finite-temperature expression \cite{Mald} for ${\rm Im}\chi$. This, however,
would only affect the
momentum relaxation rate to higher order in $T/T_{\rm F}$, and we may
therefore use the zero-temperature expression, since we are not interested
in variations on the scale of the Fermi temperature, but  on a much
smaller temperature scale set by the distance between the two
layers. As we shall see in sect.\ 4 below, this
characteristic temperature scale
is given approximately by $T_{\rm F}/(k_{\rm F}d)^{\alpha}$,
where $\alpha\simeq 0.8$.

By using the identities
\begin{equation}
\delta(\epsilon_{1} + \epsilon_{2} - \epsilon_{1'} - \epsilon_{2'})
=
\hbar\int d\omega\delta(\epsilon_{1}  - \epsilon_{1'} - \hbar\omega)
\delta(  \epsilon_{2}  - \epsilon_{2'}+\hbar \omega)
\end{equation}
and
\begin{equation}
f^0(\epsilon)(1-f^0(\epsilon +\hbar\omega))=
(f^0(\epsilon)-
f^0(\epsilon +\hbar\omega))/(1-\exp(-\hbar\omega/kT))
\end{equation}
we may transform the expression (\ref{lign1})  into
\begin{eqnarray}
\sum_{\sigma_1}^{}\int_{}^{}\frac{d{\bf k}_1}{(2\pi)^2}k_{1x}
(\frac{\partial f_1}{\partial t})_{\rm coll}&=&
-\frac{e\hbar^2 E_2\tau_2}{8 \pi^2mkT}
\sum_{\sigma_1\sigma_2\sigma_{1'}\sigma_{2'}}^{}
       \int_{}^{}\frac{d{\bf q}}{(2\pi)^2}\nonumber \\
& &
\int_{0}^{\infty} d\omega ({\rm Im} \chi({\bf q},\omega))^2
 w(1,2;1',2')
\frac{q^2}{\sinh^2 (\hbar \omega/2kT)}
\label{lign2}
\end{eqnarray}

To proceed further we must specify the effective interaction.
Our starting point is the direct Coulomb interaction between
electrons in each layer, suitably modified by
the screening in the two-dimensional electron gas.
For simplicity we assume  static screening, treated in the Thomas-Fermi
approximation, which in two dimensions yields the same
result as the random-phase approximation provided the
wave-vector transfer in the collisions between electrons is less than twice
the Fermi wave-vector. The effective interaction
is  obtained by solving Poisson's equation
for the potential due to a point source situated in one of the
two layers. When the wave-vector transfer is much larger than
both the inverse distance between the two layers and the Thomas-Fermi
screening wave-vector, the Fourier-transform $e\phi(q)$ of the
effective interaction reduces to that of the bare Coulomb interaction
between charge densities localized at the two quantum wells.
In the general case treated in App.\ A, $\phi$ is shown to be of the form
\begin{equation}
e\phi(q)=\frac{2\pi e^2}{\kappa q_{\rm TF}}\,\frac{q/q_{\rm TF}}
{g_{12}^{-1}(g_{11}+q/q_{\rm TF})^2-g_{12}}.
\label{exact}
\end{equation}
Here $g_{11}=-2qG_{11}$ and $g_{12}=-2qG_{12}$ are  functions of $q$,
where  $G_{12}$ and $G_{11}$ are form factors
given by (A.24) and (A.25),
while $q_{\rm TF}$ is the Thomas-Fermi screening wave-vector
appropriate to two dimensions,
\begin{equation}
q_{\rm TF}=\frac{2me^2}{\kappa\hbar^2},
\end{equation}
 $\kappa$ being the  dielectric constant (for  GaAs $\kappa\simeq 13$).

The simplest case to consider is that in which the width $L$ of
the quantum wells
is effectively zero, which corresponds to treating the
two layers as mathematical planes.
Then the functions $g_{11}$ and $g_{12}$ reduce to
\begin{equation}
g_{11}= 1;\;\;\;g_{12}=e^{-qd}.
\end{equation}
We may take into account the finite width of the quantum wells
by assuming a specific form of the
 electron wave function in the direction perpendicular to the
layers. If the latter is approximated by the
ground-state wave function of an infinitely deep well,
the functions
 $G_{11}(q)$ and $G_{12}(q)$ become those given by (A.24)
and (A.25).

When the collision probability is obtained
from the effective interaction by use of the
 Born approximation (or, equivalently, the golden rule), we therefore get
\begin{equation}
\sum_{\sigma_1\sigma_2\sigma_{1'}\sigma_{2'}}^{}
w(1,2;1',2')  =\frac{2\pi}{\hbar} 4|e\phi(q)|^2.
\end{equation}
The factor of four arises from doing
the spin-summations, taking into account that $w$ vanishes
when $\sigma_{1'}\neq\sigma_{1}$ and  $\sigma_{2'}\neq\sigma_{2}$.

Collecting these results  and using (\ref{2})-(\ref{4})
together with
$u_2=e\tau_2{ E}_2/m$ we can express
the ratio between the electric fields in layer 1 and 2
in terms of the momentum relaxation rate $1/\tau_D$
according to
\begin{equation}
\frac{E_1}{E_2}=\frac{\tau_2}{\tau_D}
\end{equation}
where
\begin{eqnarray}
\frac{1}{\tau_D}=
\frac{\hbar^2 }{2\pi^2n_1mkT}
       \int_{0}^{\infty}dq
\int_{0}^{\infty} d\omega q^3 |e\phi(q)|^2({\rm Im} \chi({ q},\omega))^2
\frac{1}{\sinh^2 (\hbar \omega/2kT)}.
\label{lign3}
\end{eqnarray}
Note that the factor $q^2$, which enters the
 momentum  relaxation rate under consideration, removes
the singularity at small $q$.

We have exhibited in Fig.\ 2 the $q$- and $\omega$-dependence of
the integrand in  (\ref{lign3}). Note that the integrand
vanishes for $q=0$, while it is non-zero for $\omega\rightarrow 0$,
$q\neq 0$, since ${\rm Im}\chi$ depends linearly
on $\omega$ for small values of $\omega$ \cite{komm}.

The remaining task is now to carry out the
integrations in  (\ref{lign3}), which we shall discuss
in the following section.

\newpage

4. {\sc RESULTS AND DISCUSSION}

In this concluding section we shall first consider the limit
of very low temperatures, in which an analytic result may
be extracted. Then we exhibit the results of our numerical calculation
and  compare with the experiments of \cite{gram}.

\vspace{0.5cm}

4.1  {\sc The temperature-dependent rate of momentum relaxation}

At sufficiently low temperatures we may approximate Im$\chi$
by its low-frequency expansion, Im$\chi \simeq
m^2\omega/2\pi\hbar^3qk_{\rm F}$, valid for $q\ll k_{\rm F}$.
 This allows one to complete
the integration over $\omega$, since
\begin{equation}
\int_{0}^{\infty}dx\frac{x^p}{4\sinh^2 x/2}=p!\zeta(p).
\label{tint}
\end{equation}
The remaining integral over $q$
involves the interaction matrix element $\phi(q)$,
which is determined selfconsistently in App.\ A.
For simplicity we assume for the moment that the width $L$ of the
two  layers may be set equal to zero (cf.\ Fig. 6,
in which the effect of a finite layer thickness is illustrated),
corresponding to the expression (A.7),
\begin{equation}
e\phi(q) =
\frac{2\pi e^2}{\kappa}
\frac{q}
{2q_{\rm TF}^2\sinh qd +(2qq_{\rm TF}+q^2)\exp qd}.
\label{tiln}
 \end{equation}
Since the important contributions from
the $q$-integration come from the region,
in which $q$ is comparable to or smaller than $d^{-1}$,
and since $d^{-1}$  is much
smaller than both $k_{\rm F}$ and $q_{\rm TF}$
under the experimental conditions of \cite{gram},
we may neglect the second term in the denominator of (\ref{tiln})
and thus use the simple form
\begin{equation}
e\phi(q) =
\frac{\pi e^2}{\kappa}
\frac{q}
{q_{\rm TF}^2\sinh qd }.
\label{tiln1}
\end{equation}
The final integration over
  $q$ may now be carried out, using (\ref{tint}). This  results in
\begin{equation}
\frac{1}{\tau_D}=\frac{\zeta(3)\pi}{16}
   \frac{k^2T^2}{\hbar\epsilon_{\rm F}}
\frac{1}{(k_{\rm F}d)^2}
\frac{1}{(q_{\rm TF}d)^2}.
\label{slut}
\end{equation}
Our low-temperature result (\ref{slut}) agrees with
\cite{gram}  apart from being a
factor of two larger.

The quadratic temperature-dependence given by (\ref{slut})
only applies to the low-temperature limit. At
higher temperatures the integration must be carried out numerically.
 In the following we present a set of curves that illustrate the behavior
of $1/\tau_D$ as a function of temperature and
layer separation.
  In Fig.\ 3 we plot  $1/\tau_DT^2$  as
a function of temperature for two different values
of the layer separation, $d=375$ {\AA} and $d=425$ {\AA},
 with the parameter $L$ describing our model wave functions
taken to be $L=200 $ {\AA},
appropriate to the experiments reported in \cite{gram}.
Note that  $1/\tau_DT^2$ at high temperatures decreases
with increasing temperature. By
 performing a high-temperature expansion
(with the restriction that $T\ll T_{\rm F}$) of
(\ref{lign3}),
using $1/\sinh^2(\hbar\omega/2kT)\simeq 4k^2T^2/\hbar^2\omega^2
-1/3$, one finds  that $1/\tau_{D}\propto T - {\rm const.}/T$,
which explains the observed fall-off at higher temperatures \cite{gram}.

 In Fig.\ 4 the dependence
of $d^4/\tau_DT^2$ on $T$ is plotted
for three different values of $d$. The curves illustrate
 that the momentum relaxation rate is  proportional to
$d^{-4}$ only in the low-temperature limit.     Note
that the temperature $T_{\rm max}$, at which the curves
have their maximum, shifts to lower temperatures when $d$ is increased.
In the $T \rightarrow 0$ limit, the analytic formula (\ref{slut}) gives
$\tau_D^{-1} d^4/T^2 \rightarrow 1.38 \cdot
 10^{-24} {\rm m}^4/{\rm K}^2{\rm s}$,
which is slightly larger
 than the extrapolated value from Fig.\ 4.  This result
can be understood by examining Fig.\ 6, where one observes that the
approximate effective interaction (\ref{tiln1}) is
 slightly larger than the full
expression (\ref{exact}) used in the numerical calculation of Fig.\ 4.

Our numerical calculation shows that
$T_{\rm max}$ is approximately proportional to
$d^{-\alpha}$, where $\alpha$ is close to
0.8. This power-law behavior is illustrated
 in Fig.\ 5 for two different choices of the parameter $L$.
The exponents obtained from the fit to a power law
are  nearly the same: $\alpha= 0.81$  and $\alpha = 0.76$ for
$L=200$ {\AA} and $L=1$ {\AA}, respectively.
Finally we show in Fig.\ 6  how $|\phi(q)|^2$
depends on $q$, for  different forms
of the effective interaction  considered in App. A.

\vspace{0.5cm}

4.2  {\sc Comparison with experiment}

The experimental data obtained in  \cite{gram} have
a form similar to our theoretical curves in Fig.\ 3, but the
temperature at which
$1/\tau_DT^2$ has its maximum is significantly less, about 2.5 K
in the experiment of \cite{gram}. The Fermi temperature
corresponding to the experimental value of
the number of electrons per unit area is 62 K. The maximum of
the theoretical curves occurs around  10 K  for the values of $d$
shown in Fig.\ 3. Furthermore the observed rate is
about a factor of two
larger than our calculated
values, suggesting that other mechanisms of momentum transfer may be
important.

\newpage

5.  {\sc Conclusion}

In this paper we have considered the rate of momentum transfer
between two nearby electron gases, which are coupled via
the screened Coulomb interaction. This system is of
particular interest,
because it provides a novel way of probing electron-electron
interactions in two dimensions, which play a crucial role
in many problems of current interest. It is well known that
going from three to two dimensions may change qualitatively
predictions based on phase-space arguments, the classical
example being the logarithmic energy, or temperature,
dependence of the electron life-time. These effects
can be related mathematically to logarithmic singularities in the
phase-space integral, which determines the life-time, or scattering rate.
The phase-space corresponding to the momentum transfer rate
is different from the phase space related to the life-time:
the singularity corresponding to small momentum transfer is
removed, and the $2k_{\rm F}$-singularity is strongly
suppressed due to the effective interaction, which decays exponentially
for large momentum transfer. Thus it might seem that
there is no reason to expect deviations from a $T^2$-temperature
dependence. It was therefore a significant surprise, when recent measurements
revealed a non-monotonic temperature dependence of the
momentum transfer rate (when scaled by $T^2)$.  To account quantitatively
for the experimental observations, it is likely that other mechanisms
than the direct Coulomb coupling must be considered.
However, as we have shown in this paper, the Coulomb coupling
 itself yields a nontrivial temperature dependence, which
bears a clear resemblance to  the one observed experimentally.

Finally we wish to stress that the system under study
is very versatile, since a number of external parameters can
 be adjusted. As an example, in this work we have considered
only the case of two  Fermi circles of equal radius. The phase-space
considerations are quite different for the case of two different
 Fermi momenta, and it is therefore of interest to generalize our analysis
to this case, and to perform measurements on two unequal quantum wells.

\vspace{0.5cm}

{\bf Acknowledgment}

We thank J. P. Eisenstein for correspondance.

\newpage

\setcounter{page}{19}

$^{\dagger}$ Also at Microelectronics Centre,
Technical University of Denmark, 2800 Lyngby, Denmark.


\begin{thebibliography}{99}
\bibitem{komm} The slight irregularities seen on Fig.2 are an artefact of the
graphics routine,
which we employ,
 and in reality the integrand is smooth. The sharp ridge requires
a large number of mesh points in the numerical integration, and
typically 140000 function evaluations are needed to
reach convergence.
\end{thebibliography}
\end{document}